\documentclass[twocolumn,showpacs,amsmath,amssymb,floatfix,prl,aps,superscriptaddress,showkeys]{revtex4}

\usepackage{graphicx}
\usepackage{epsfig}
\usepackage{dcolumn}
\usepackage{bm}
\usepackage{times}
\usepackage{color}
\usepackage{soul}

\begin{document}
\title{Exact partition function for the random walk of an electrostatic field}
\author{Gabriel Gonz\'alez}\email{gabriel.gonzalez@uaslp.mx}
\affiliation{C\'atedras CONACYT, Universidad Aut\'onoma de San Luis Potos\'i, San Luis Potos\'i, 78000 MEXICO}
\affiliation{Coordinaci\'on para la Innovaci\'on y la Aplicaci\'on de la Ciencia y la Tecnolog\'ia, Universidad Aut\'onoma de San Luis Potos\'i,San Luis Potos\'i, 78000 MEXICO}
\begin{abstract}
The partition function for the random walk of an electrostatic field produced by several static parallel infinite charged planes in which the charge distribution could be either $\pm\sigma$ is obtained. We find the electrostatic energy of the system and show that it can be analyzed through generalized Dyck paths. The relation between the electrostatic field and generalized Dyck paths allows us to sum over all possible electrostatic field configurations and is used for obtaining the partition function of the system. We illustrate our results with one example.
\end{abstract}

\maketitle

\section{I. Introduction}
\label{sec1}
In 1961 A. Lenard consider the problem of a system of infinite charged planes freely to move in one direction without any inhibition of free crossing over each other \cite{lenard}. If all the charged planes carry a surface mass density of magnitude unity, and the $i^{th}$ one carries an electric surface charge density $\sigma_i$, the Hamiltonian of Lenard's problem is given by
\begin{equation}
{\mathcal H}(q_1,q_2,\ldots;p_1,p_2,\ldots)=\frac{1}{2}\sum_{i}p_i^2-\frac{1}{2\epsilon_0}\sum\sum_{i<j}\sigma_i\sigma_j|q_i-q_j|,
\label{eq01}
\end{equation}
where the system is overall neutral. With the Hamiltonian function at hand Lenard was able to obtain the exact statistical mechanis of the system with the technique of generating functions \cite{lenard,lenard1}. \\
From a physical point of view, Lenard's model represents a one-dimensional Coulomb gas \cite{bax}. The one dimensional Coulomb gas is a statistical mechanical problem where particles of equal or opposite charges interact through the Coulomb potential \cite{prager}. The model has been extensively studied in the past and forms one of the classical exactly soluble problems in one dimension \cite{elliot}. Also, the Coulomb gas model has also been used to describe the main features of ionic liquids \cite{dem,dem1}.\\
More recently, a very similar problem was proposed in Ref. \cite{gg} with the constraint that each of the charged planes has a fixed position in space and that the surface charge distribution in each plane could be either $\pm\sigma$. This modified model gives rise to a random walk behavior of the electrostatic field that can be analyzed as a Markovian stochastic process.\\
The purpose of this paper is to give a statistical description of the electrostatic field generated by several static parallel infinite charged planes in which the surface charge distribution could be either $\pm\sigma$. The key step in our formulation consists of the summation over all possible trajectories of the electrostatic field for every different charge configuration. We do this by showing that there is a one to one correspondence between every electrostatic field trajectory and a generalized Dyck path.\\
The article is organized as follows. In section (II) we describe the model and derive a system of equations for obtaining the electrostatic energy of the system or Hamiltonian. In section (III) we make a one to one correspondence between electrostatic field trajectories and generalized Dyck paths. In section (IV) the explicit expression for the partition function is given in terms of generating functions. In the last section we summarize our conclusions.\\

\section{II. Model of the system}
\label{sec2}
Suppose that we are given a collection of $2N$ infinite charged planes, half of them have a constant charge distribution $\sigma$ and the other half have a constant charge distribution $-\sigma$, i.e. the system is overall neutral. If we randomly place the $2N$ infinite charged planes parallel to each other along the $z$ axis at position $z_n = nd$ where $n\in \{0,1,\dots,2N-1\}$, then the electrostatic field would evolve along the $z$ axis making random jumps each time it crosses an infinite charged sheet.
\begin{figure}[!hb]
			\centering
			\includegraphics[width=0.3\textwidth]{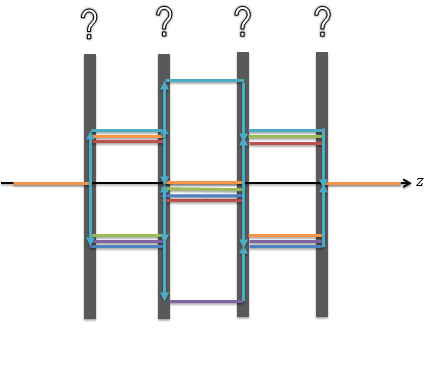}
			\caption{Possible electrostatic field configurations inside a four charged plane system where the charge distribution is not known.}
			\label{fig:field1}
\end{figure}
For example, consider a system of four charged planes. We know that $E(z)$ equals zero at the left and right side of the configuration due to the neutrality of  the system \cite{Griffiths}. After crossing the first charged plane, $E(z)$ would increase or decrease in an amount of $\sigma/\epsilon_0$ depending if the first charged plane had a positive or negative surface charge density. In figure \ref{fig:field1} we show all the possible electrostatic field configurations for the case when the charge density is not known.\\
Since we are concern with the statistical properties of the electrostatic field we have first to obtain the Hamiltonian of the field or the electrostatic energy of the system. We know from basic electrodynamics courses that the electrostatic energy is given by \cite{Griffiths} 
\begin{equation}
{\mathcal E}=\delta_{E}\Delta V
\label{eq02}
\end{equation}	
where $\delta_E=\epsilon_0 E^2/2$ is the electrostatic energy density and $\delta V=Ad$ is the volume inside the region between two consecutive parallel charged planes. Denoting $E(i)$ as the value of the electrostatic field in the $i^{th}$ region between two charged planes placed in $z=(i-1)d$ and $z=id$, then we have the following boundary condition:
\begin{equation}
E(i)-E(i-1)=\frac{\sigma}{\epsilon_0}S_i, \quad \mbox{ for $i=1,2,\ldots,2N$ },
\label{eq03}
\end{equation}
where $S_i$ is a random variable which takes one of the two possible values $\pm 1$ depending if the charged plane at $z~=(i-1)d$ is positively or negatively charged. \\
We can write down the boundary condition given in Eq.(\ref{eq03}) in matrix form in the following way
\begin{equation}
\left[\begin{array}{ccccc}
1 & 0 & 0 & 0 & 0 \\
-1& 1 & 0 & 0 & 0 \\
0 & -1& 1 & 0 & 0 \\
\vdots & & \ddots & \ddots & \vdots \\
0 & \cdots & & -1 & 1\\
\end{array}\right]\left[\begin{array}{c}
E(0)\\ E(1)\\ E(2)\\ \vdots \\ E(2N)\\ \end{array}\right]=\frac{\sigma}{\epsilon_0}
\left[\begin{array}{c} 
0\\ S_1 \\ S_2 \\ \vdots \\ S_{2N}\end{array}\right],
\label{eq04}
\end{equation}
where we have included the initial condition of the electrostatic field, i.e. $E(0)=0$. Solving Eq. (\ref{eq04}) for the electrostatic field we have
\begin{equation}
E(i)=\frac{\sigma}{\epsilon_0}\sum_{j=1}^{i}S_j, \quad \mbox{for $i=1,2\ldots,2N$.}
\label{eq05}
\end{equation}
Substituting Eq. (\ref{eq05}) into Eq. (\ref{eq02}) we obtain the electrostatic energy of the system which is given by
\begin{equation}
{\mathcal E}=\left(\frac{\epsilon_0}{2}Ad\sum_{i}E^2(i)\right)\delta_{\sum_{i=1}^{2N}S_i,0},
\label{eq06}
\end{equation}
where the Kronecker delta guarantees the neutrality of the system.
\section{III. Electrostatic field and Dyck paths}
\label{sec3}
\begin{figure}[!hb]
			\centering
			\includegraphics[width=0.4\textwidth]{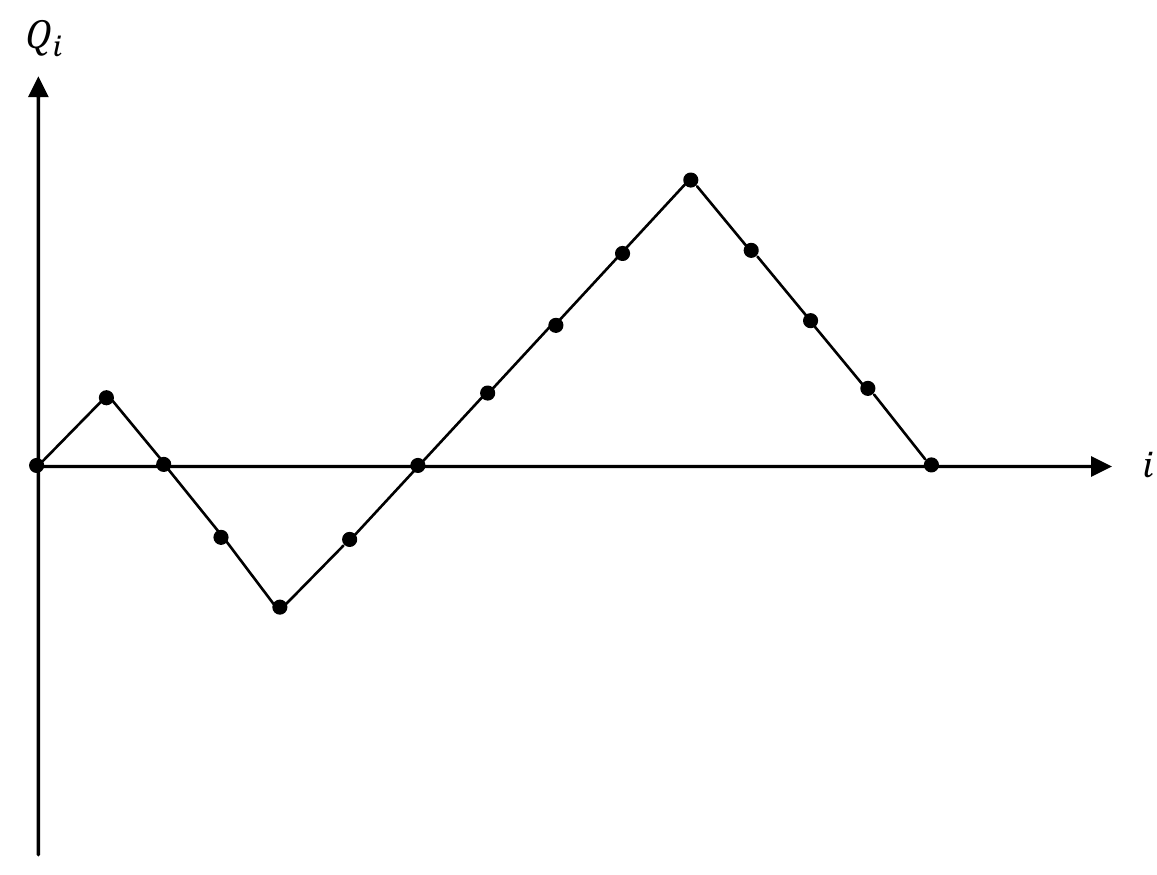}
			\caption{Generalized Dyck paths representing a possible electrostatic field configuration of $14$ parallel static charged planes.}
			\label{dycke}
\end{figure}

In this section we present the relation between the electrostatic field and generalized Dyck paths. Let us consider for the sake of simplicity that all the charged planes carry a surface charge density of $\sigma=\pm 1/A$. Let $Q_i$ be the total amount of charge to the left of the $i^{th}$ region, then we can rewrite the electrostatic field given in Eq. (\ref{eq05}) as
\begin{equation}
E(i)=\frac{Q_i}{\epsilon_0A}.
\label{eq07}
\end{equation}
Note that $Q_0=Q_{2N}=0$ and $|Q_i-Q_{i-1}|=1$. The electrostatic energy in terms of the new variable is given then by
\begin{equation}
{\mathcal E}=\frac{1}{2C}\sum_{i}Q_i^2
\label{eq08}
\end{equation}
where $C=\epsilon_0A/d$.\\
It is convenient to investigate the nature of all the possible electrostatic field configurations by plotting $E(i)$. For the sake of simplicity we will graph $Q_i$ versus $i$ and connect all the different values of the electrostatic field in each region by straight lines. An example of a graph is shown in Fig. (\ref{dycke}). 
Note that every possible electrostatic field configuration consists of a closed path of length $2N$ that starts at $(0,0)$ and ends at $(2N,0)$, and is uniquely determined by a sequence of vertices $(Q_0,Q_1,\ldots,Q_{2N})$, where the $Q_{j-1}$ is connected to $Q_{j}$ by an edge \cite{gopal,dent}. In the mathematical literature this is known as a closed {\it simple walk} of $2N$ steps on the lattice of integers, where the discrete time that numbers the steps is measured in the horizontal axis, and the positions of the walker on the line are recorded in the vertical axis. Since our walks start from the origin, at time $j$ the walker is on site $s(j)=Q_0+Q_1+\ldots+Q_{j}$. A Dyck path is a simple closed walk with the following constraint $s(j)\geq 0$ for all $j$ \cite{chapman}. The constraint on a Dyck path means that it never falls below the horizontal axis, but it may touch it at any number of integer points. A more general Dyck path or generalized Dyck paths are simple closed walks that are allowed to go below the horizontal axis \cite{jac}. We note then that all the electrostatic field configurations are the ensembles of generalized Dyck paths of length $2N$ that start at $(0,0)$ and end at $(2N,0)$. \\
Let us denote by $\Gamma$ the ensemble of generalized Dyck paths of length $2N$ and consider $\gamma\in\Gamma$, then the partition function of the electrostatic field will be given by \cite{cicuta}
\begin{equation}
{\mathcal Z}=\sum_{\gamma\in\Gamma}e^{-\beta{\mathcal E}(\gamma)}
\label{eq09}
\end{equation} 
where $\beta=1/k_BT$ and $k_B=1.38\times10^{-23}$J/K is Boltzmann constant and $T$ is the temperature. Equation (\ref{eq09}) is the same as calculating a Feynman amplitude in the frame of generalized Dyck paths, i.e. one sums over all generalized Dyck paths of length $2N$ and to each whole path is then attributed a weight which is given by the Boltzmann factor. 
\section{IV. Exact partition function}
\label{sec4}
In order to calculate the sum of Eq. (\ref{eq09}) we need to count over all possible generalized Dyck paths of length $2N$. A very powerful tool for counting walks is by means of the transfer matrix method \cite{shapiro,jon}. A transfer matrix method can be used to count all possible generalized Dyck paths of length $2N$. Let us introduce a set of $(2N+1)\times(2N+1)$ matrices of the form $\langle i|W(\ell)|j\rangle$ which represent the number of simple walks from $i$ to $j$ in $\ell$ steps. Since the paths we are considering are made of steps $\pm 1$, then the transfer matrix for this case is a bidiagonal one given by
\begin{equation}
W_{ij}=\left[\begin{array}{cccccc}
0 & b_{\text{\tiny{-N,-N+1}}} & 0  & 0& 0 &0 \\
a_{\text{\tiny{-N+1,-N}}}& 0 & b_{\text{\tiny{-N+1,-N+2}}}  &0 &0 & 0 \\
0 & a_{\text{\tiny{-N+2,-N+1}}}& 0  & \ddots & 0 & \vdots\\
0 & 0 & \ddots &\ddots & b_{\text{\tiny{N-1,N-1}}} & 0 \\
\vdots &\vdots & 0 & a_{\text{\tiny{N-1,N-2}}}& 0 &b_{\text{\tiny{N-1,N}}}  \\
0 & \cdots &  0 & 0 & a_{\text{\tiny{N,N-1}}} & 0\\
\end{array}\right].
\label{eq10}
\end{equation}
where
\begin{equation}
a_{\text{i+1,i}}=e^{-\frac{\beta i^2}{2C}},\,\,  b_{\text{i,i+1}}=e^{-\frac{\beta(i+1)^2}{2C}}\,\,\, \mbox{for $-N\leq i\leq N$}
\label{eq11}
\end{equation}
In this picture, the evaluation of $W_{ij}^{\ell}$ consists in summing the weights of all walks of $\ell$ steps from $i$ to $j$. Therefore, the partition function for the random walk of an electrostatic field generated by means of $2N$ static parallel randomly charged planes where half of them have a charge surface density $\sigma=1/A$ and the other half has a charge surface density $\sigma=-1/A$ is given by
\begin{equation}
{\mathcal Z}=[W_{i,j}^{2N};N+1,N+1],
\label{eq12}
\end{equation} 
where $[W_{i,j}^{2N};N+1,N+1]$ represents the element in row $N+1$ and column $N+1$ of the transfer matrix $W_{ij}$ raised to the $2N$ power.\\
For example, consider the case of 4 parallel static randomly charged planes with surface charge distribution $\sigma=\pm 1/A$, for this case $N=2$ and we can then write down the transfer matrix for this particular problem which is
\begin{equation}
W_{ij}=\left[\begin{array}{ccccc}
0 & e^{-\beta/2C} & 0 & 0 & 0 \\
e^{-4\beta/2C}& 0 & 1 & 0 & 0 \\
0 & e^{-\beta/2C}& 0 & e^{-\beta/2C} & 0 \\
0 & 0& 1 & 0 & e^{-4\beta/2C} \\
0 & 0 &0 & e^{-\beta/2C} & 0\\
\end{array}\right]
\label{eq13}
\end{equation}
therefore, the partition function will be given by 
\begin{equation}
{\mathcal Z}=[W_{i,j}^{4};3,3]=2e^{-3\beta/C}+4e^{-\beta/C}.
\label{eq14}
\end{equation}
Equation (\ref{eq14}) is given in the form ${\mathcal Z}=\sum_{k}g({\mathcal E}_k)e^{-\beta{\mathcal E}_k}$, where $g({\mathcal E}_k)$ represents the number of paths that share the same energy ${\mathcal E}_k$, this means that for the case when $T\rightarrow\infty$ all the possible electrostatic field configurations will have the same probability distribution, i.e. 
\begin{equation}
\lim_{T\rightarrow\infty}P({\mathcal E}_k)=\lim_{T\rightarrow\infty}\frac{e^{-\beta{\mathcal E}_k}}{{\mathcal Z}}=\frac{(N!)^2}{(2N)!},
\label{eq15}
\end{equation}
this was the case studied in Ref. \cite{gg}.\\
We can calculate the expectation value for the energy of the system when $T\rightarrow 0$, i.e.
\begin{equation}
\lim_{T\rightarrow 0}\langle {\mathcal E}\rangle=\lim_{T\rightarrow 0}\left(-\frac{\partial \ln{\mathcal Z}}{\partial\beta}\right)=\frac{N}{2C}.
\label{eq16}
\end{equation} 
Equation (\ref{eq16}) means that, for very low temperatures, only the paths with minimum electrostatic energy will contribute to the energy of the system. For this case, the system behaves like $N$ parallel plate capacitors connected in series.
\section{V. Conclusions} 
We have shown that it is possible to obtain the exact partition function for the electrostatic field generated by means of several static parallel infinite charged planes in which the surface charge distribution is not explicitly known. We have worked out the special case where the charged planes have a constant surface charge distribution given by $\pm 1/A$ and the overall electrostatic system is neutral, i.e. there is the same number of positive and negative charged planes. We use the connection between generalized Dyck paths and possible electrostatic field configurations to obtain the partition function of the system as a discrete Feynman path integral. A transfer matrix approach was used to count over all possible generalized Dyck paths.
\section{Acknowledgments}
This work was supported by the program ``C\'atedras CONACYT".

\end{document}